\documentstyle[twocolumn,epsbox]{jpsj}

\def\mbf(#1){\mbox{\boldmath $#1$}}
\def\runtitle{Proximity effects near the interface 
}
\def\runauthor{Kazuhiro {\sc Kuboki} }

\title
{Proximity Effects near the Interface between $d$-wave Superconductors 
and Ferro/Antiferromagnets
}

\author{Kazuhiro KUBOKI}

\inst{Department of Physics, Kobe University, Kobe 657-8501, Japan}

\recdate{ \hskip3.0truecm }

\abst{  
We study the proximity effects near the interface between a 
$d$-wave superconductor (S) and a ferromagnet (F)  or an 
antiferromagnet (AF).  
The S-side (F and AF-side) is described by the attractive (repulsive) 
Hubbard model, and the Bogoliubov de Gennes equation 
derived within the Hartree-Fock approximation is solved numerically. 
The superconducting order parameter and the magnetization ($m$) can 
coexist near the interface, and the spatial variation of $m$  induces 
the spin-triplet $p$-wave component in both F and AF cases. 
The local density of states is also calculated and discussed. 
}

\kword{unconventional superconductivity, proximity effect, 
Bogoliubov de Gennes equation, local density of states, 
$SO(5)$ theory 
}

\begin{document}
\sloppy
\maketitle
Recently the proximity effect of unconventional superconductors has 
been a subject of intensive study.\cite{Demler,YT,add6}
This is because the interface properties of these superconductors 
can be quite different from those of conventional ($s$-wave) 
ones due to the nontrivial angular structure of pair wave 
functions, so that their study is of partiucular interest.   
Now many unconventional superconductors are known to exist,  
and the most famous examples are the 
high-$T_c$ cuprates in which the $d_{x^2-y^2}$-wave superconducting (SC) 
state is realized. \cite{Scal}

In this letter we study the proximity effect between a 
$d$-wave superconductor (S) and a ferromagnet (F)  or an 
antiferromagnet (AF). 
We examine the possible coexistence of magnetism and 
superconductivity near the interface. 
In $s$-wave superconductors the coexistence is unlikely or limited 
because of the full gap in the excitation spectrum, while in the $d$-wave 
case it is not. We will show that the coupling of SC order parameters (OPs)
with the magnetization induces a spin-triplet $p$-wave component. 
We also calculate the local density of states (LDOS) 
to show how the electronic structure changes across the interface. 

The system we consider is a two-dimensional 
$d$-wave superconducto(S)/magnet(M) bilayer. 
The direction perpendicular (parallel) to the 
interface is denoted as $x$ ($y$), and we assume that the system 
is uniform along the $y$-direction except the two-sublattice 
spin structure in the AF case. 
We treat the Hubbard model on a square lattice within the 
Hartree-Fock (HF)  approximation to describe the magnetism and 
superconductivity at zero temperature ($T = 0$). 
The crystal $a$-axis ia taken to be parallel to the $x$-direction,  
i.e., we consider only the [100] interface. 
The Hamiltonian for the two layers is given by 
\begin{equation} 
\begin{array}{rl}
  H_L & =  - \displaystyle t_L \sum_{<i,j> \sigma} 
    ( c_{i,\sigma}^{\dagger} 
      c_{j,\sigma} + h.c.)  
     +  \sum_i U_L n_{i\uparrow}n_{i\downarrow}  \\ 
     + & \displaystyle 
     \sum_{<i,j>} V_L \big[n_{i\uparrow}n_{i\downarrow}  
     + n_{j\uparrow}n_{i\downarrow}  \big], 
     \ \ (L = M, S) 
\end{array}     
\end{equation}
where $\langle i,j \rangle $ and $\sigma$ denote the nearest-neighbor 
pairs and the spin index, respectively. 
Parameters $t_L$, $U_L$ and $V_L$ are the 
transfer integral, the on-site interaction and the nearest-neighbor
interactions, respectively, for the $L (=M, S)$ side.  
The transmission of electrons at the interface is described by the 
following tight-binding Hamiltonian:
\begin{equation}
    H_T =  - \displaystyle t_T \sum_{<l,m>\sigma} 
    ( c_{l,\sigma}^{\dagger} 
      c_{m,\sigma} + h.c.)     
\end{equation}
where $l$ ($m$) denotes the surface sites of $M$ ($S$) layer, 
and then the total Hamiltonian of the system is $H = H_M + H_S + H_T 
- \mu \sum_{i\sigma} c_{i\sigma}^\dagger c_{i\sigma}$ with 
$\mu$ being the chemical potential.  
We have examined the various values of $t_T/t_{M(S)}$. 
The results are qualitatively similar,  and the effect of proximity 
is reduced for smaller $t_T$, as expected. 
Thus the discussion is restricted to the case of $t_T = t_M = t_S \equiv t$ 
throughout this letter. 

The interaction terms are decoupled within the HF approximation
\begin{equation}
\begin{array} {rl}
U n_{i\uparrow}n_{i\downarrow} \to &
U\langle n_{i\uparrow} \rangle n_{i\downarrow} 
+ U\langle n_{i\downarrow} \rangle n_{i\uparrow}
-U \langle n_{i\uparrow} \rangle \langle n_{i\downarrow} \rangle, \\
Vn_{i\uparrow}n_{j\downarrow} \to &
V\Delta_{ij}c_{j\downarrow}^\dagger c_{i\uparrow}^\dagger 
+V\Delta_{ij}^{*}c_{i\uparrow} c_{j\downarrow}
-V\vert \Delta_{ij} \vert^2 
\end{array}
\end{equation}
with $\Delta_{ij} \equiv \langle c_{i\uparrow}c_{j\downarrow}\rangle$, 
and $\Delta_{ij}$ and magnetization 
$m_i =\langle n_{i\uparrow} - n_{i\downarrow} \rangle/2$ are the OPs 
to be determined self-consistently. 
Along the $y$-direction we introduce two-sublattices $A$ and $B$, 
and carry out the Fourier transformation: 
$c_{i\sigma}^\alpha= \sqrt{2/N_y}\sum_k 
c_{x_i,\sigma}^\alpha(k) e^{iky_i}  \ \ (\alpha = A, B) $,  where $N_y$ 
($N_x$) is the number of sites along the $y$ ($x$) direction. 
Now we define 
$\Delta_{ij}^{\alpha} = 
\langle c_{i\uparrow}^\alpha c_{j\downarrow}^\alpha \rangle$
($\alpha = A, B$),  
$\Delta_y^{(1)}(x_i) = 
\langle c^A_{i\uparrow}c^B_{i+{\hat y}\downarrow} \rangle$ and 
$\Delta_y^{(2)}(x_i) = 
\langle c^A_{i\downarrow}c^B_{i+{\hat y}\uparrow} \rangle$. 
Then the mean-field Hamiltonian is written as (hereafter  
$x_i$ is abbreviated as $i$) 
\begin{equation}
{\cal H}_{\rm MFA} = \sum_k\sum_i\sum_j \Psi_i^\dagger(k) 
{\hat h}_{ij}(k) \Psi_j(k)
\end{equation} 
where 
\begin{equation}
{\hat h}_{ij}=
\left [\begin{array}{clcr}
W^A_{ij\uparrow}(k) & F^A_{ij}(k) &  {\tilde W}_{ij}(k) & F_{ij}(k) \\
F^A_{ij}(k)^{*} & -W^A_{ij\downarrow}(k) & F_{ij}^{*}(k) & -{\tilde W}_{ij}(k) \\
{\tilde W}_{ij}(k) & F_{ij}(k)  & W^B_{ij\uparrow}(k)   &  F^B_{ij}(k) \\
F_{ij}^{*}(k)  & -{\tilde W}_{ij}(k) & F^B_{ij}(k)^{*} &  -W^B_{ij\downarrow}(k)
\end{array}\right ]
\end{equation} 
\begin{equation} 
\displaystyle \Psi_i^\dagger(k) \equiv [c^A_{i\uparrow}(k)^\dagger,   
c^A_{i\downarrow}(-k), c^B_{i\uparrow}(k)^\dagger, 
c^B_{i\downarrow}(-k)] .
\end{equation}
with
\begin{equation}
\begin{array}{rl}
W^\alpha_{ij\sigma}= &\displaystyle 
-t(\delta_{j,i+1}+\delta_{j,i-1}) 
+ (U\langle n^\alpha_{i,-\sigma}\rangle -\mu)\delta_{ij} \\
{\tilde W}_{ij}(k) = & \displaystyle -2t \cos k \ \delta_{ij} \\
F_{ij}^\alpha = & -V\Delta^\alpha_{ij}
(\delta_{j,i+1}+\delta_{j,i-1}) \\
{\tilde F}_{ij}(k) = & \displaystyle - \frac{V}{2}
\big[e^{ik}\Delta_y^{(1)}(i) - e^{-ik}\Delta_y^{(2)}(i)\big]
\delta_{ij} . 
\end{array}
\end{equation}

We diagonalize the mean-field Hamiltonian by solving the following
Bogoliubov de Gennes (BdG) equation:\cite{dG}
\begin{equation} 
\sum_j {\hat h}_{ij}(k) u_{jn}(k) = E_n(k) u_{in}(k) , 
\end{equation} 
where $E_n(k) $ and $u_{in}(k)$ are the energy eigenvalue and the 
corresponding eigenfunction, respectively, for each $k$. 
The unitary transformation $\Psi_i(k) = \sum_n u_{in}(k) \Gamma_n(k)$ 
diagonalizes the matrix ${\cal H}_{\rm MFA}$, and  
conversely the OPs $\Delta_{ij}$ and $m_i$ 
can be written in terms of $E_n(k) $ and $u_{in}(k)$. 
These constitute the self-consistency equations which  
will be solved numerically in the following.\cite{Nishi}

In a uniform (bulk) case, the ground-state phase diagram of the repulsive 
Hubbard model ($U > 0$ and $V = 0$)
within the HF approximation was examined by 
Hirsh.\cite{Hirsh,Penn}
There the ferro-, antiferro- and paramagnetic states are obtained 
depending on the value of $U/t$ and the electron density. 
For an attractive case ($V < 0$ and $U = 0$), the ground state is a spin-singlet 
SC state. 
Depending on the electron density a $d_{x^2-y^2}$- ($\Delta_d$) or an 
extended $s$-wave ($\Delta_s$)-wave SC state is stabilized, 
and the former is favored near half-filling. 
Here, $\Delta_d$ ($\Delta_s$) can be constructed as a linear 
combination of $\Delta_{ij}$'s in such a way that it changes its sign 
(is invariant) under 90$^\circ$ rotation in the basal plane. 

Now, let us study the M/S bilayer system. 
We impose the open (periodic) boundary condition for the 
$x$ ($y$) direction, 
and the typical system size treated is $N_x \times N_y = 
40 \times 40$ to  
$60 \times 80$ sites. 
We use $t$ as a unit of energy (i.e., $t = 1$).  
First we choose $U_M > 0$, $V_M = 0$,    
$U_S = 0$ and $V_S < 0$
so that the ferromagnetic and $d$-wave 
superconducting states are stabilized in M and S layers, respectively 
(F/S system).\cite{finite} 
In Fig.1 the spatial variations of OPs are shown.
It is seen that the magnetization ($m$) and the  
SCOP ($\Delta$) coexist near the interface,  
and the $p_x$-wave component ($\Delta_{px}$) as well as $\Delta_s$ appears. 
In general, the OP component different from that in the bulk can be 
induced due to the scattering of Cooper pairs at 
an interface (or a surface faced to vacuum).
The important point here is that we get 
the $p_x$-wave state which is a spin-triplet (even parity) SC state.
This state has pairing OP on the bonds along the $x$-direction. 
The spatial variations of OPs in the AF/S bilayer system are 
shown in Fig.2. In this case the spin-triplet $p_y$-wave component 
($\Delta_{py}$) also appears.  We will neglect the small $\Delta_s$ in 
the following discussions. 

The penetration of $m$ into the S side causes the 
imbalance of the densities of spin-up and spin-down electrons.
Thus electron pairs cannot be formed in singlet channels 
only, and then the spin-triplet component appears. 
It is seen that the M side is essentially unaffected by the proximity effect.
This is because the Curie (or N\'eel) temperature is higher than 
$T_{\rm c}$ of superconductivity so that the typical lengths for the decay 
of the induced OPs are shorter in the M side. 
Then in the following we will mainly focus on the S side. 

We analyze the above results more precicely using the 
Ginzburg-Landau (GL) theory. 
(The GL theory is not quantitatively valid except near $T_{\rm c}$, 
but it can give qualitatively correct results.) 
The GL free energy in the S layer (in the continuum representation) 
can be written as\cite{SigUe}
\begin{equation} 
\begin{array}{rl}
{\cal F_S} = &\displaystyle \int d^2r \Bigl(\sum_{j=d,s,px,py} 
\big[\alpha_j|\Delta_j|^2 + K_j|\partial  \Delta_j|^2\big] \\
+ & \displaystyle K_{ds} \big[(\partial_x \Delta_d)(\partial_x \Delta_s)^* 
- (\partial_y \Delta_d)(\partial_y \Delta_s)^* + c.c.\big] \\
+ & \displaystyle K_{dp} \big[\Delta_d \big\{(\partial_xm_z)\Delta_{px}^*-
(\partial_ym_z)\Delta_{py}^*\big\}m_z  + c.c\big] \\
+ & \displaystyle K_{sp} \big[\Delta_s \big\{(\partial_xm_z)\Delta_{px}^*+
(\partial_ym_z)\Delta_{py}^*\big\}m_z + c.c\big]\Bigr). 
\end{array}
\end{equation}
This ${\cal F}_S$  is invariant under all symmetry  operations for the square 
lattice and we have dropped higher order terms.
All coefficients in ${\cal F}_S$ are positive definite 
except $\alpha_j$ which are given at a temperature $T$ as,  
$\alpha_j =  V_S 
\Bigl[1 -  (V_S/N) \sum_k w_j^2(k) 
\tanh(\xi_k/2T)/2\xi_k \Bigr] $
for $j = d, s $ and 
$\alpha_j =  (V_S/2) 
\Bigl[1 -  (V_S/N) \sum_k w_j^2(k) 
\tanh(\xi_k/2T)/\xi_k \Bigr] $
for $j = p_x, p_y$.   Here $N = N_x N_y$, 
$\xi_k = -2t(\cos k_x + \cos k_y) -\mu$,  
$w_d(k) = \cos k_x - \cos k_y$, 
$w_s(k) = \cos k_x +  \cos k_y$ and 
$w_{px(y)} = \sin k_{x(y)}$. 
In the bulk superconductor,  the SCOP with the highest $T_{\rm c}$ 
occurs ($T_{\rm c}$ is given by $\alpha(T_{\rm c}) = 0$) and other 
components are usually suppressed due to the 
higher order repulsive coupling to the bulk OP 
(not shown in ${\cal F}_S$).\cite{SigUe}
Near half-filling $w_d(k)^2$ ($w_s(k)^2$) is large (small) on the Fermi 
surface, and $\alpha_d$ becomes negative first. 
This is the reason we obtain the $d$-wave state near half-filling. 
(On the contrary,  an extended $s$-wave state is favored away from 
half-filling.)

In the M/S bilayer system, several OPs can coexist 
and they are not spatially uniform. 
In the F/S case, $\partial_x m \not= 0$, $\partial_y m = 0$ and 
$\Delta_d \not= 0$, so that $\Delta_{px}$ is induced but 
$\Delta_{py}$ is not. On the contrary, in the AF/S system,  finite 
$\partial_y m$ induces $\Delta_{py}$ also. 
The opposite sign of $\Delta_{px}^A$ and $\Delta_{px}^B$ and their 
oscillating behavior in AF/S system can also be understood as 
being due to the staggered nature of $m$.
The phases of the induced OPs relative to the bulk one ($\Delta_d$) 
are either 0 or $\pi$, since they are determined by the bilinear coupling 
terms in ${\cal F}_S$. 
Namely, the state has the ($d_{x^2-y^2} \pm p_x$)-symmetry near the 
interface of the F/S system, 
and the ($d_{x^2-y^2} \pm p_x \pm p_y$)-symmetry in the 
AF/S case. 

It has been proposed that antiferromagnetism and $d$-wave 
superconductivity can be treated in a unified way using 
$SO(5)$ symmetry.\cite{Zhang,Mura}  
The present result seems consistent with this theory, since $m$ 
penetrates deeper into the S side in the AF/S case than in 
the F/S case. 
On the contrary $\Delta_{px}$ decays faster in the former. 
The reason for this is the following. 
The F/S system studied here is closer to half-filling 
compared with AF/S system.\cite{renorm}
Then, $\alpha_{px}$ in ${\cal F}_S$ can be closer to zero 
(though still positive)  in the F/S case because the factor $w_{px}^2$ 
is larger on the Fermi surface. 
The coherence length, which determines the decay of $\Delta_j$,    
is given by $\xi_j =\sqrt{K_j/\alpha_j}$, and then the decay is slower 
in the F/S case. ($K_j$ is not so sensitive to the electron density.) 

Recently, Honerkamp et al.\cite{Honer} studied the surface states 
of $d$-wave superconductors.
They found that near the [110] surface (faced to vacuum) 
the staggered magnetization can appear spontaneously,  and that 
the $p$-wave as well as the extended $s$-wave component can occur. 
Their state breaks time-reversal symmetry spontaneously 
and a surface current is 
generated.\cite{TBrk,SigR,KK,TBrk2,Sauls,add3,add2,add4,add5} 
The reason the $p$-wave OP is obtained in the present 
case is, however,  different from theirs. 
We consider a [100] surface and if it is faced to vacuum, 
the magnetization cannot be spontaneously generated, hence no 
$\Delta_p$ is obtained.  
In order to have a  finite $\Delta_p$ for the case of a  [100] surface, 
the proximity effects from the magnets are necessary. 
Honerkamp et al. found the bound states near the Fermi level, which 
are different for the spin-up and spin-down components because of 
local antiferromagnetism. 
If the [110]-oriented M/S bilayer is treated, the split 
bound states would also be expected. 

Next, we show the results of the local density of 
states (LDOS). \cite{add1}
The LDOS at site $i (\in A)$ is given by 
\begin{equation}
\begin{array}{rl}
N^A_\uparrow (i, \omega) = & \displaystyle \frac{2}{N_y}
\sum_k \sum_n |u_{4i-3,n}(k)|^2 \delta(\omega-E_n(k)) \\
N^A_\downarrow (i, \omega) = & \displaystyle \frac{2}{N_y}
\sum_k \sum_n |u_{4i-2,n}(k)|^2 \delta(\omega+E_n(k)) 
\end{array}
\end{equation}
($\uparrow, \downarrow$ 
being the spin indices) and $N^B_\uparrow(i, \omega)$ 
($N^B_\downarrow(i, \omega)$) is 
obtained by replacing $(4i-3)$ ($(4i-2)$) in the above equation 
with $(4i-1)$, ($(4i)$). 
For the F/S case $N^A_\sigma = N^B_\sigma$, and for the AF/S system
$N^A_\sigma = N^B_{-\sigma}$ 
($\sigma= \uparrow, \downarrow$).  
In figures 3 and 4 (5 and 6), 
the results for the F/S (AF/S) system are shown. 
Deep inside the F (AF) layer, the LDOS for spin-up 
and the spin-down components is split as a result of magnetic 
orders (Fig.3 (Fig.5)). 
The spin-up and spin-down components gradually merge 
as the interface is crossed, and a $V$-shaped DOS which is 
typical of a $d$-wave superonductor is formed. 
These changes of the LDOS as a function of the position
appear smooth. 

In summary we studied the proximity effects near the interface 
between $d$-wave superconductors and ferro/antiferromagnets. 
We found the coexistence of the magnetic order and the superconductivity,  
which leads to the ($d_{x^2-y^2}\pm p_x$)-wave and 
($d_{x^2-y^2}\pm p_x\pm p_y$)-wave surface states in the F/S  and 
AF/S systems,  respectively. 
The LDOS changes smoothly from magnetic to SC layers. 
In this letter we did not take into account the effect of a vector potential 
that couples to the magnetization in the ferromagnetic state. 
This coupling may lead to more drastic effects, and is most suitably 
treated by field theoretical methods. 
Such an effective theory can be constructed based on the knowledge 
of the present solutions of BdG equations.  
This problem will be examined separately in the near future.

\smallskip
The author thanks M. Sigrist for many useful discussions. 
This work was supported in part by a Grant-in-Aid for Scientific 
Research (C) from the Ministry of Education, Science, Sports and Culture 
of Japan.



\bigskip
 \noindent 
  {\bf Fig. 1}  
  Spatial variations of OPs for a F/S bilayer. 
  Here $U_M = 10, U_S = 0, V_ M = 0, V_S = -2$ and $\mu=-0.1$. 
  Note that all OPs are non-dimensional.
  
  \noindent 
  {\bf Fig. 2}
   Spatial variations of OPs for an AF/S bilayer. 
  Here $U_M = 4, U_S = 0, V_ M = 0, V_S = -2$ and $\mu=1$. 
  (a) $\Delta_d$, $\Delta_s$ and $m_{A(B)}$. 
  (b) $\Delta_{px}^{A(B)}$, $\Delta_{py}$ and $\Delta_d$. 
  Indices $A$ and $B$ denote the sublattice. 

  \noindent 
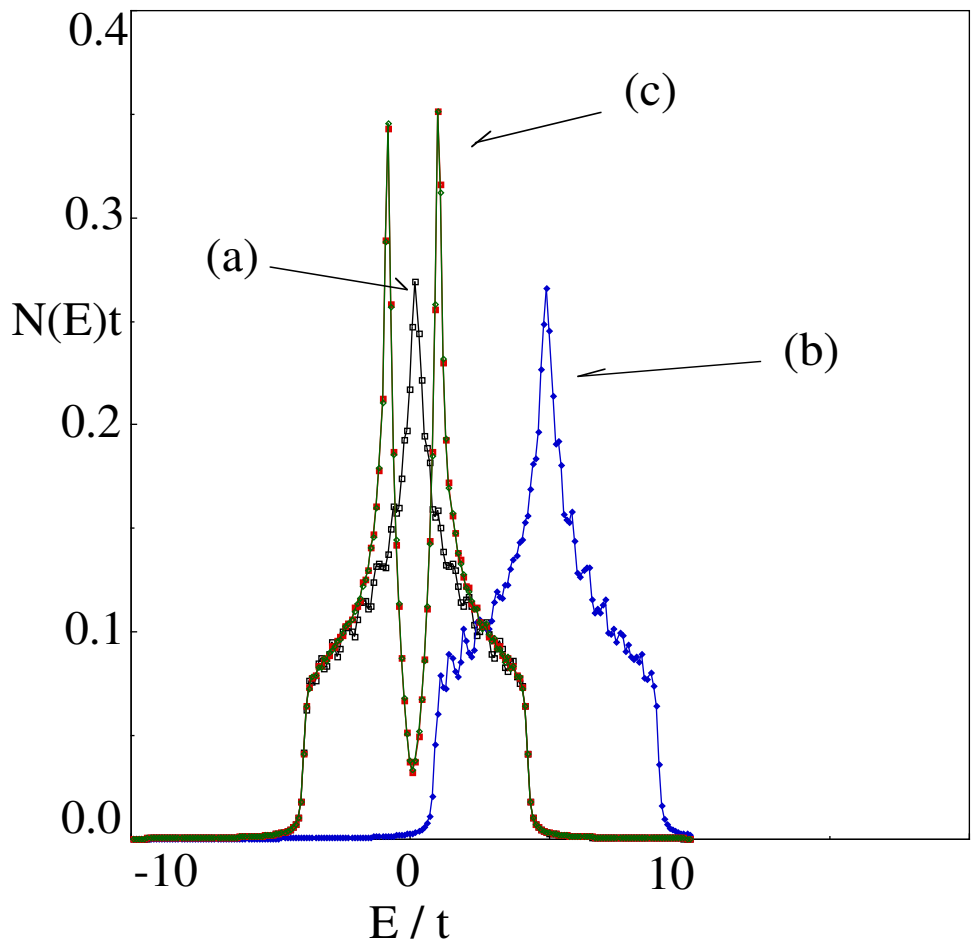
  {\bf Fig. 3} 
  LDOS for a F/S bilayer. 
  The parameters are the same as in Fig.1 and the finite width 
  $\Gamma=0.08$ is introduced to each state. 
  (a)  $x = -9a$ with $\sigma = \uparrow$; 
  (b)  $x = -9a$ with $\sigma = \downarrow$, 
  (c) $x = 21a$ ($\sigma =\uparrow$ and $\downarrow$ are almost 
  degenerate). Here, $a$ is the lattice constant and $x = 0$ 
  corresponds to the surface site of the M layer. 
  
  \noindent
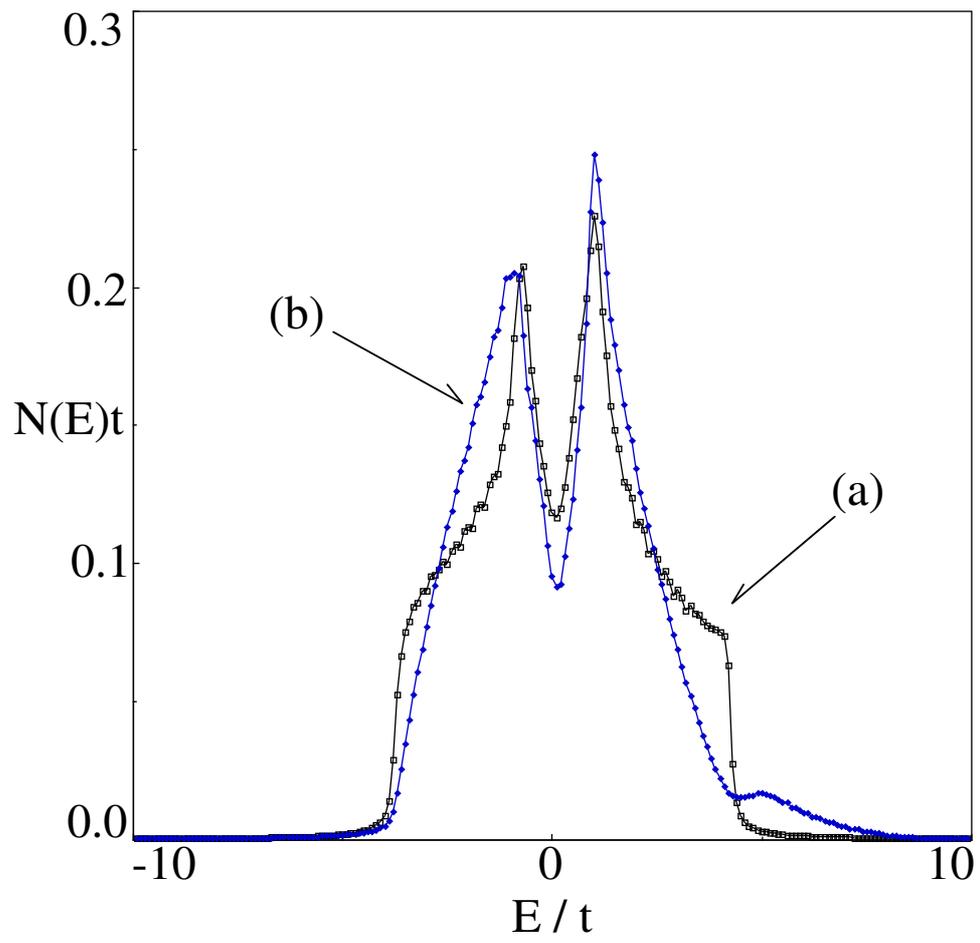
  {\bf Fig.4} 
  LDOS for a F/S bilayer. The parameters are the same as in Fig.3 except 
  $x = a$.  (a) $\sigma = \uparrow$,  (b) $\sigma = \downarrow$.
  
  \noindent 
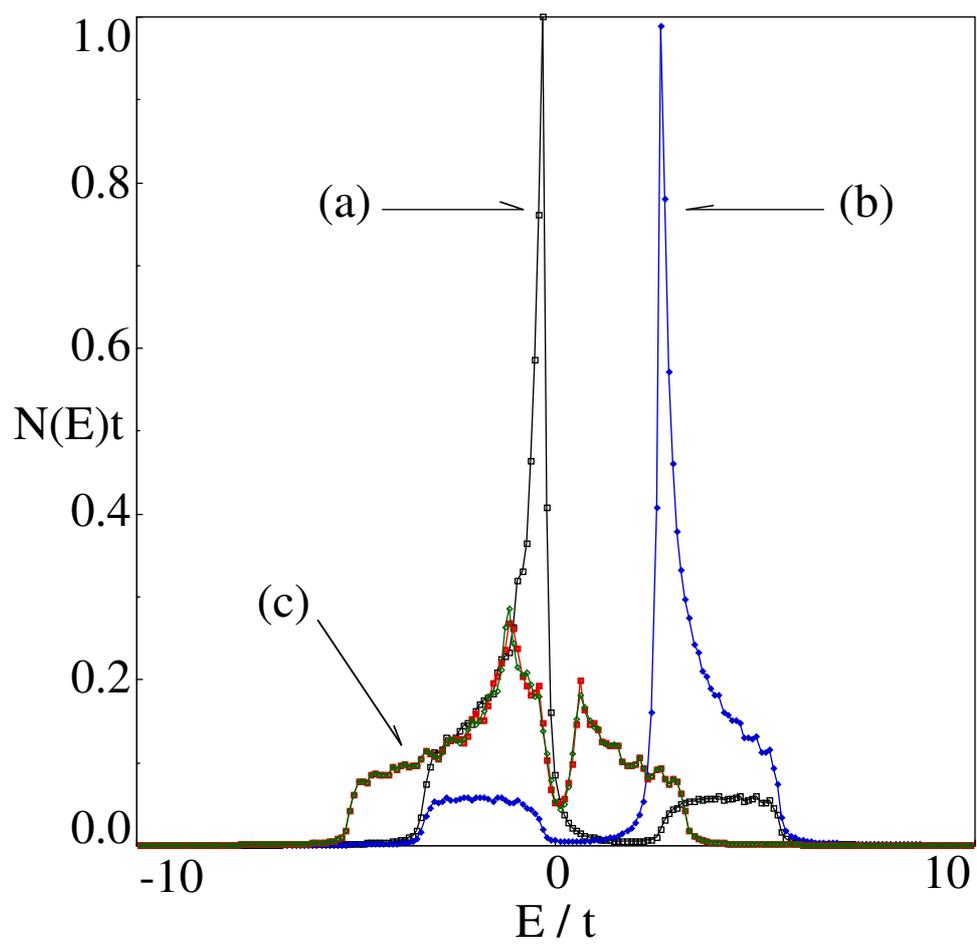
  {\bf Fig. 5}   
  LDOS for an AF/S bilayer. 
  The parameters are the same as in Fig.2 and the finite width 
  $\Gamma=0.08$ is introduced.  
  (a) $x = -9a$ with $\sigma = \uparrow$; 
  (b) $x = -9a$ with $\sigma = \downarrow$, 
  (c) $x = 11a$ ($\sigma =\uparrow$ and $\downarrow$ are almost 
  degenerate). 
  
  \noindent 
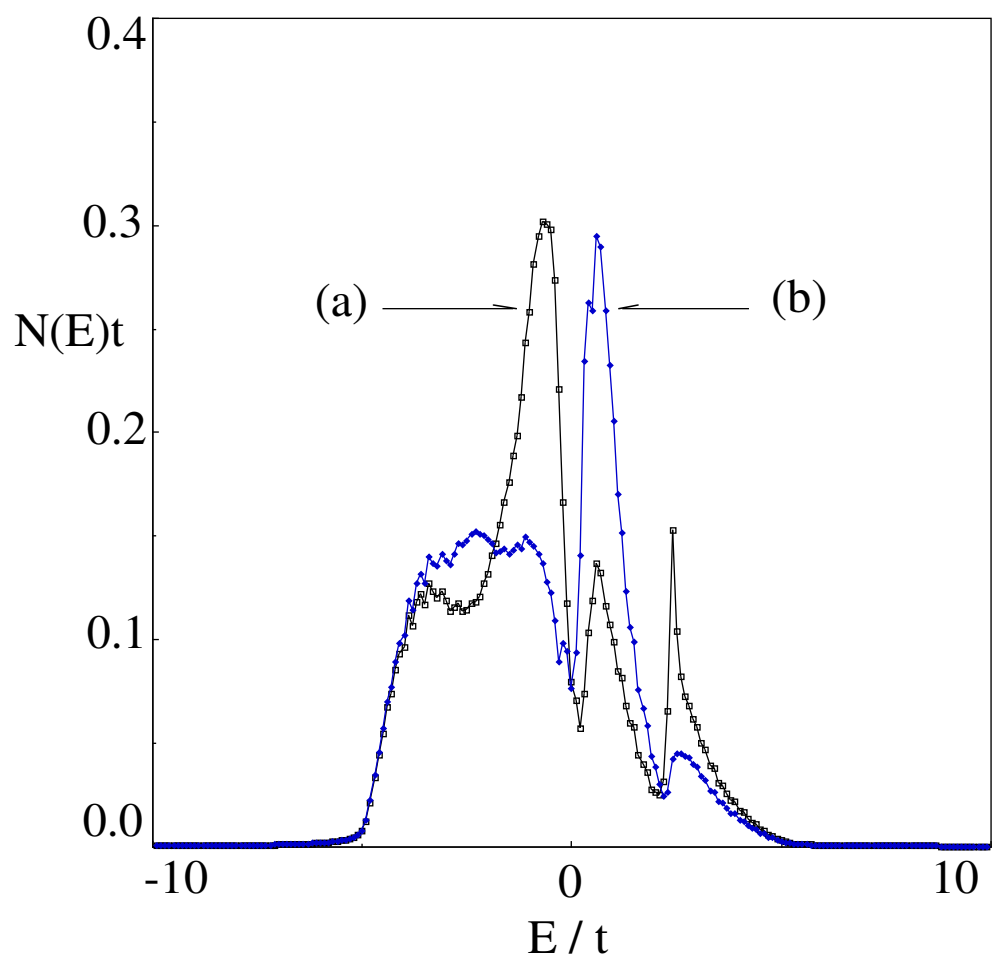
  {\bf Fig.6} 
   LDOS for an AF/S bilayer. The parameters are the same as in Fig.5 
   except $x = a$. (a) $\sigma = \uparrow$,  (b) $\sigma = \downarrow$.
  

\begin{thebibliography}{99}
\bibitem{Nishi} For numerical diagonalization of matrices we have used 
the computer program written by H. Nishimori: TITPACK version2.
\bibitem{finite} We have also studied the case of finite $U_S$ and 
$V_M$ ($0 < U_S \ll U_M, |V_M| \ll |V_S|$), but the results were 
essentially the same.
\bibitem{renorm} Although bare chemical potential $\mu$ is close to 0, 
there is a renormalization of $\mu$ from the Hartree term for $U$ 
in the F side.
\end{thebibliography}
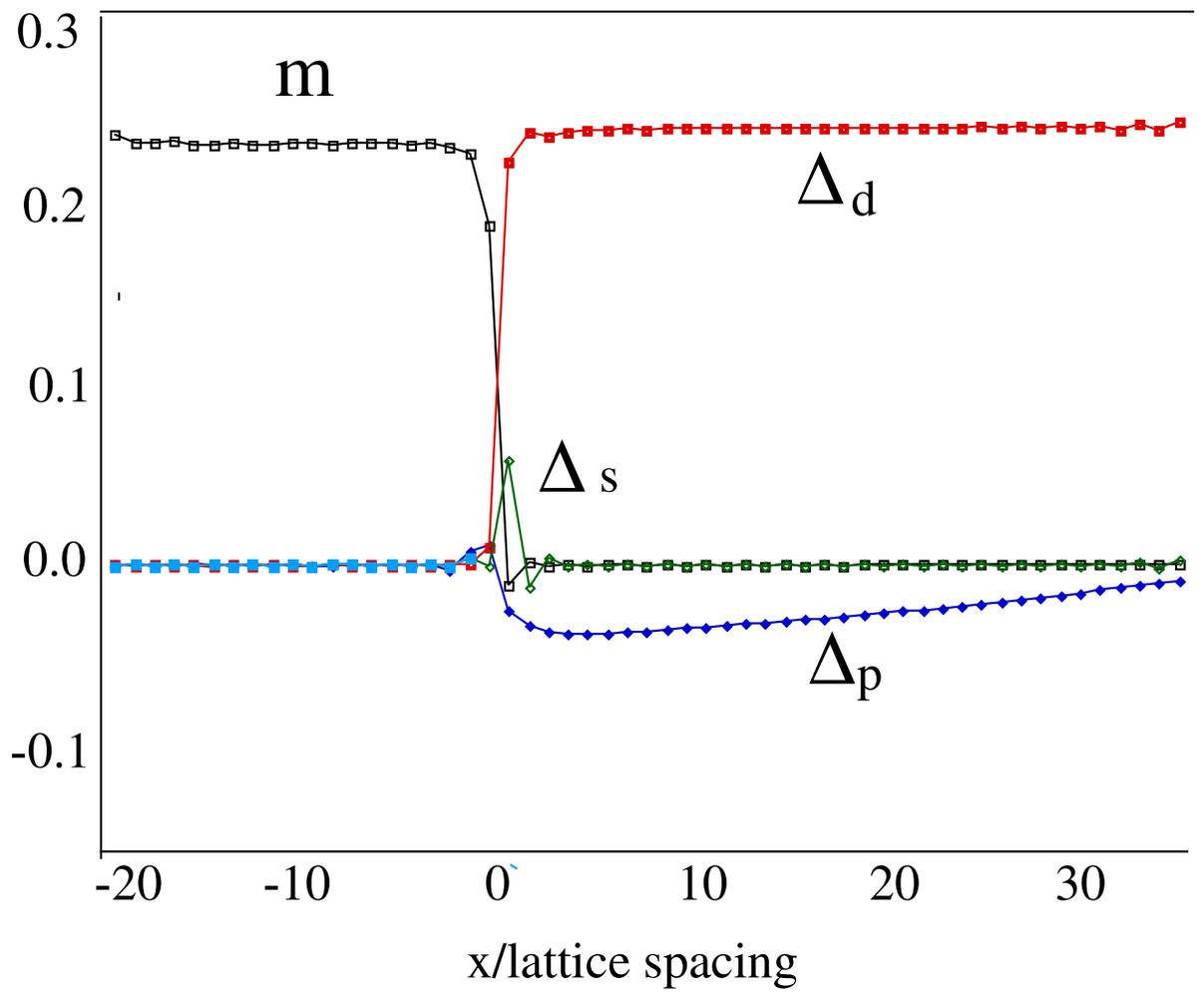
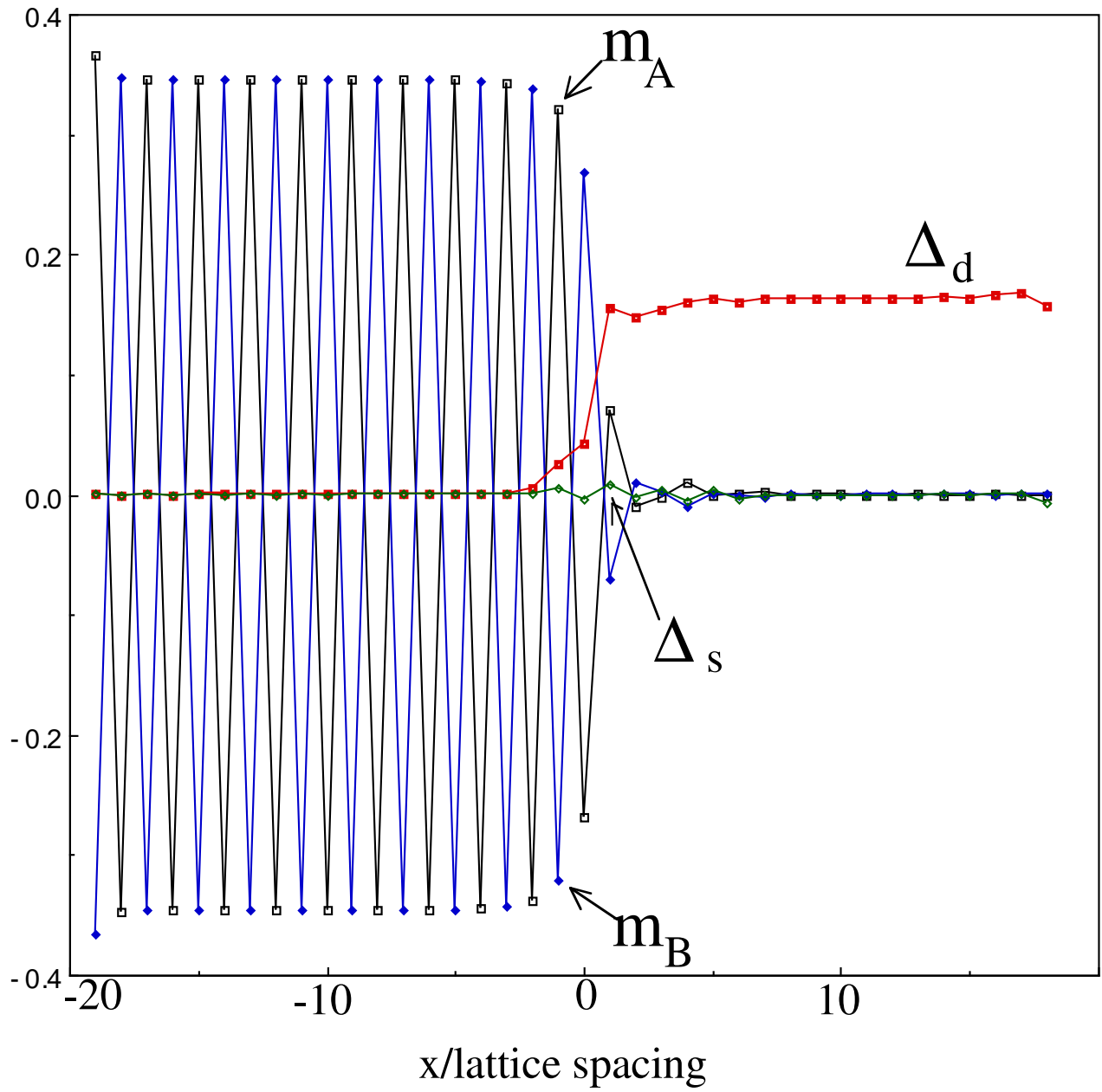
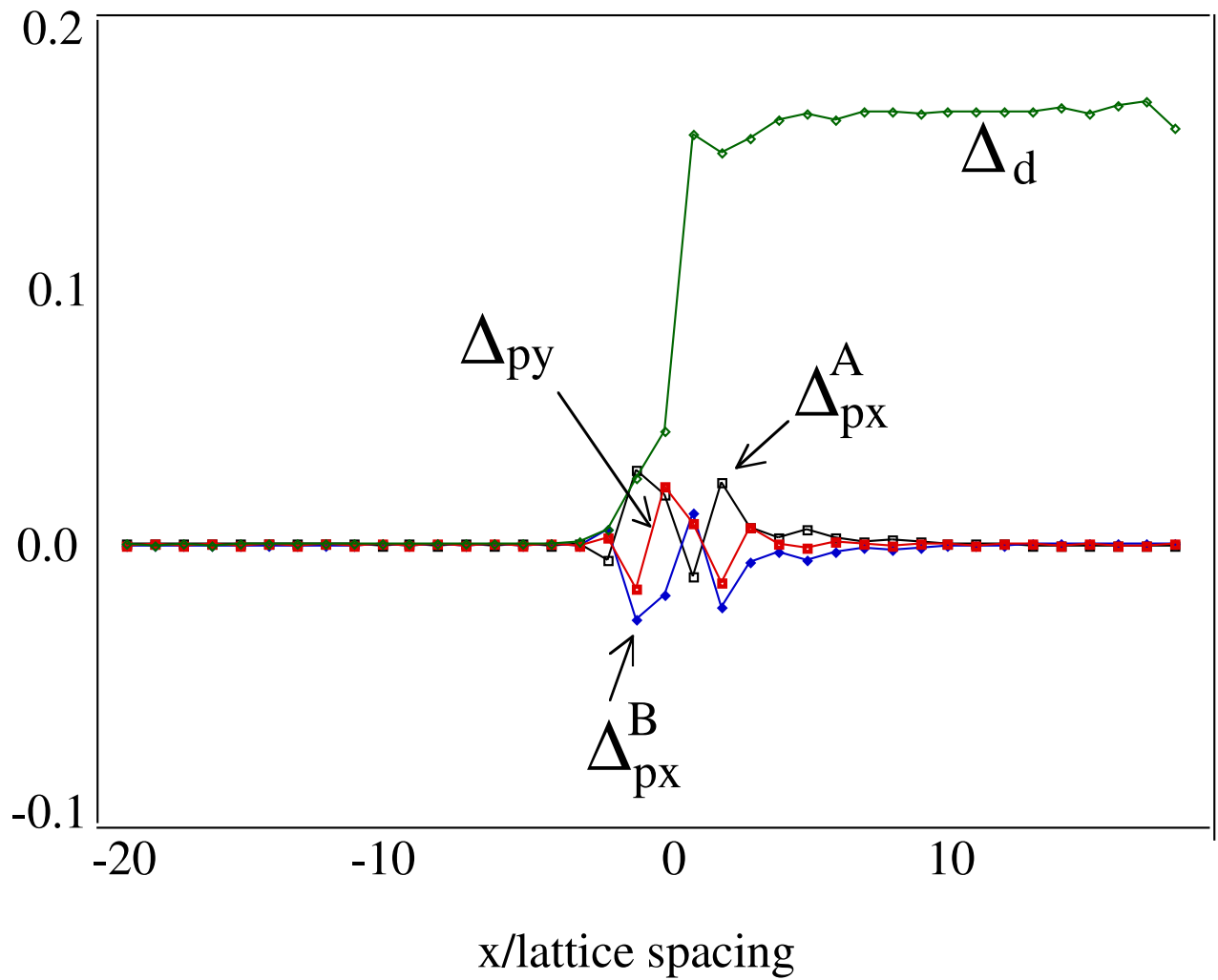
\end{document}